\documentclass[12pt,a4paper]{article}

  \usepackage{a4wide}
  \usepackage{latexsym}
  \usepackage{epsf}
  \usepackage{amssymb}
  \usepackage{graphicx}
  \usepackage{amsmath, cite}
  \usepackage{amsmath,amssymb,amsthm}
  \usepackage{verbatim}
  \usepackage{xcolor}
  \usepackage[colorlinks=true,linkcolor=black,citecolor=black,urlcolor=blue,linktocpage,pageanchor=false,debug]{hyperref}
  \usepackage{cancel}

\renewcommand{\d}{\textrm{d}}
\newcommand{\e}{\textrm{e}}

\newcommand{\be}{\begin{equation}}
\newcommand{\ee}{\end{equation}}

\newcommand{\ads}{\mathrm{AdS}_7 \times M_3}

\begin{document}
\numberwithin{equation}{section}

\begin{center}

\begin{flushright}
{\small UUITP-19/13} \normalsize
\end{flushright}
\vspace{0.7 cm}

{\LARGE \bf{A note on smeared branes in flux vacua \\ \vspace{0.5cm} and gauged supergravity  }}

\vspace{1 cm} {\large  U. H. Danielsson$^a$, G. Dibitetto$^a$, M. Fazzi$^b$,  T. Van
Riet$^c$} \footnote{\href{mailto:giuseppe.dibitetto@physics.uu.se}{\ttfamily giuseppe.dibitetto}, \href{mailto:ulf.danielsson@physics.uu.se}{\ttfamily ulf.danielsson @ physics.uu.se}, \href{mailto:mfazzi@ulb.ac.be}{\ttfamily \mbox{mfazzi @ ulb.ac.be}}, \href{mailto:thomasvr@itf.fys.kuleuven.be}{\ttfamily \mbox{thomasvr @ itf.fys.kuleuven.be}}}\\

\vspace{0.75 cm}{$^{a}$ Institutionen f{\"o}r fysik och astronomi,
Uppsala Universitet,\\ Box 803,   SE-751 08 Uppsala, Sweden}\\

\vspace{0.2 cm}  \vspace{.15 cm} {$^b$ Universit\'e Libre de Bruxelles and International Solvay Institutes,\\ ULB-Campus Plaine CP231, B-1050 Brussels, Belgium}

\vspace{0.2 cm}  \vspace{.15 cm} {$^c$ Instituut voor Theoretische Fysica, K.U. Leuven,\\
Celestijnenlaan 200D, B-3001 Leuven, Belgium}

\vspace{1.2cm}

{\bf Abstract}
\end{center}


{\small In the known examples of flux vacua with calibrated spacetime-filling sources (branes or orientifold planes), one can smear the source in order to perform a standard KK reduction and obtain a lower-dimensional supergravity description. Furthermore, it is expected that the smeared and localized solution preserve equal amounts of supersymmetry.  In this note we point out that the $\mathrm{AdS}_7$ solution discussed in arXiv:\href{http://arxiv.org/abs/1111.2605v2}{1111.2605} and arXiv:\href{http://arxiv.org/abs/1309.2949}{1309.2949} is a counterexample to this common lore. The solution is supersymmetric when the spacetime-filling D6-branes are localized but breaks supersymmetry in the smeared limit. By using the embedding tensor formalism we demonstrate that there is no gauged supergravity description for the solution, regardless of the source being smeared or not. We conjecture that for flux solutions with separation between the KK scale and AdS radius this cannot occur. } 


\newpage
\tableofcontents
\newpage

\section{Introduction}

Despite the long history of study on flux compactifications, many interesting questions remain. Of particular interest are questions related to supersymmetry breaking and the construction of trustworthy lower-dimensional effective descriptions of flux vacua. In this paper we consider the problem of constructing effective actions for tree-level flux compactifications that involve spacetime-filling sources, such as D-branes and O-planes. A motivation for such setups is phenomenology, because orientifold planes seem necessary ingredients for the construction of flux vacua that are genuinely lower-dimensional \cite{Tsimpis:2012tu, Petrini:2013ika} (in the sense that there is a separation between the KK scale and the vacuum energy) and, at tree-level, orientifolds are necessary for having Minkowski or de Sitter vacua \cite{deWit:1986xg, Maldacena:2000mw}.  

The presence of such sources necessarily induces a warp factor in front of the lower-dimensional metric. This is in contrast with ordinary KK reduction where all dependence on internal coordinates is neglected. Nonetheless the warping can affect the low-energy physics, most notably it can soften the hierarchy problem \cite{Giddings:2001yu}. Hence we are naturally led to investigate how ordinary KK reduction is extended to warped compactifications. This can be called ``Warped Effective Field Theory'' (WEFT), see for instance \cite{Shiu:2008ry, Frey:2008xw, Martucci:2009sf, Frey:2013bha}. Typically, questions about WEFT are asked in the context of compactifications to $\mathcal{N}=1, D=4$ Minkowski vacua, with the standard example being O$3$/O$7$ compactifications on conformal Calabi-Yau spaces with three-form fluxes \cite{Giddings:2001yu,Dasgupta:1999ss}. When warping is neglected there is a standard procedure to write down the K\"ahler potential and superpotential that defines the  $\mathcal{N}=1, D=4$ supergravity that is supposed to capture the low energy physics of fluctuations around the vacuum. Technically speaking, the absence of warping\footnote{By warping we imply everything that is sourced by the O-planes and D-branes, such as a dilaton that depends on internal coordinates, the warping in front of the four-dimensional metric and the conformal factor for the internal metric, a nonzero profile for the RR form that couples to the brane.} implies that one solves the ten-dimensional equations of motion for which the sources are smeared, i.e. the delta function is replaced with a constant \cite{Grana:2006kf, Blaback:2010sj, Blaback:2012mu}. This is in spirit of ordinary KK reduction, where fields are Fourier expanded on the internal compact space and only the zero mode is kept, since zero modes have the smallest mass. However, if the warping is relevant at low energies it implies that higher order Fourier modes have low enough masses to be physically relevant and ordinary KK reduction needs to be revised.

If the supersymmetry-breaking scale is below the KK scale one expects the Wilsonian effective action (the WEFT) to be supergravity. Hence this must imply that the low energy effective theory can still be written in terms of a K\"ahler and a superpotential, but now they will get corrected by warping terms. Hence we expect two supergravity theories to exist that relate to the same flux compactification: the one obtained from smearing the source (which is an ordinary KK reduction) and the WEFT in which warping is somehow taken into account. 
The motivation for this work is that this seems problematic for theories with extended supersymmetry since such theories are usually very constrained. In minimal supergravity one could indeed think that warping corrects the K\"ahler and superpotential, but in the case of maximal or half-maximal supergravity the gauge group almost completely fixes the theory. One explanation for this could be that compactifications which preserve supersymmetry have a restricted topology and hence amount of orientifold tension. Since the tension controls the size of the warping it could be that the warping corrections are not relevant at low energies. 

With this in mind we consider a particular compactification of massive IIA supergravity to $\mathrm{AdS}_7$ space. Besides the Romans mass the other ingredients are spacetime-filling D6 branes and $H$ flux filling the internal space. In the smeared limit this solution was first found in \cite{Blaback:2010sj} where the internal space was found to be an $S^3$.  The stability with respect to the left-invariant moduli was verified in \cite{Blaback:2011nz}. The stability of the solution was not believed to be guaranteed because it was claimed to be a non-supersymmetric solution. The question of whether a sensible localized solution exists was studied in \cite{Blaback:2011nz, Blaback:2011pn} where it was shown that the localized solution must have three-form flux divergences identical to the infamous ones encountered uplifting anti-D3-branes \cite{McGuirk:2009xx, Bena:2009xk, Bena:2012bk, Gautason:2013zw}.\footnote{See \cite{Junghans:2013xza} for an overview and \cite{Blaback:2012nf} for an interpretation of the singularity.} Recently a very interesting twist to the story was given in reference \cite{Apruzzi:2013yva}: the localized solution  was found to preserve half of the supersymmetry of the ten-dimensional theory and because of that a first order integration was found that simplified the numerical study of the solution and the understanding of its global properties.  

This raises a few questions: `Does the smeared solution preserve supersymmetry?'; `Is there a $D=7$ supergravity describing the fluctuations around the vacuum (both for the smeared and localized solution)?'. The answers we find in this paper are twice negative and to our knowledge this is the first example where these phenomena occur, namely: 1) a flux vacuum that is only found to be supersymmetric when the sources are properly localized and 2) despite the very high amount of preserved supersymmetry there is no lower-dimensional supergravity description.  In the discussion we give some clues as to why this happens and when it is expected to happen. 

The rest of this paper is organized as follows. In section \ref{AdS7} we review the construction of both the smeared and localized solution and show that the smeared solution indeed breaks supersymmetry, which is rather straightforward by relying upon the results of \cite{Apruzzi:2013yva}. In section \ref{gauged} we describe the half-maximal and maximal supergravities in $D=7$ using the embedding tensor formalism \cite{deWit:2002vt, Schon:2006kz}. By constructing the dictionary between geometric fluxes and the embedding tensor components we can rule out the existence of a seven-dimensional gauged supergravity that has the aforementioned $\mathrm{AdS}_7$ vacuum as its ground state.\footnote{If there is nonzero Romans mass. If the Romans mass vanishes the solution has a lift to eleven-dimensional supergravity and the $\mathrm{AdS}_7$ vacuum can be understood as the standard Freund-Rubin vacuum describing the near horizon limit of an M5-brane.}  We conclude with a discussion in section \ref{Discussion} in which we speculate about the meaning of our results. We have included two appendices, the first of which cointains some technical details
concerning the group-theoretical calculation to derive the dictionary embedding tensor/fluxes, whereas in the second appendix the example of the no-scale Minkowski vacuum is worked out explicitly. In this example the smearing gives the gauged supergravity description and the localization is understood.

\section{$\mathrm{AdS}_7$ vacua in massive IIA supergravity}\label{AdS7}

We first review some of the results of \cite{Apruzzi:2013yva}, where all supersymmetric $\ads$ solutions obtainable from type II supergravity were found. Solutions exist only in (massive) IIA supergravity and are supported by $H$ flux filling the internal space and spacetime-filling D6-branes whose backreaction also switches on a nontrivial profile for the dilaton and the $F_2$ flux. When the Romans mass is put to zero the resulting solution (presenting D6-branes and anti-D6-branes at the poles of $M_3$, which is topologically an $S^3$) can be lifted to the well known $\mathrm{AdS}_7 \times S^4$ Freund-Rubin solution of eleven-dimensional supergravity. Here, we will only address the features of the solutions that are necessary for this paper. The $\mathrm{AdS}_7$ vacua are intriguing for many other issues, such as the appearance of diverging $H$ flux and the possible appearance of D8 stacks (that carry D6 charge).

The local properties around the $H$ flux singularities of the massive $\mathrm{AdS}_7$ solution with D6-branes of \cite{Apruzzi:2013yva} had appeared earlier in \cite{Blaback:2011nz, Blaback:2011pn}. The solutions of the latter references have extra integration constants. However, $\mathrm{AdS}_7$ solutions with extra integration constants are not expected to be globally well-defined, but they can serve as local solutions to which some geometry can be glued.\footnote{In the noncompact limit, this is the usual procedure for which the solutions are a warped product of seven-dimensional Minkowski space times conformal $\mathbb{R}^3$. It is in this limit that the connection with supersymmetry-breaking branes can be made and that the appearance of extra integrations constants is physical.} The globally well-defined solutions are supersymmetric and have no integration constants. As it turns out the solutions do break supersymmetry when the D6 sources are smeared. In the smeared limit the internal space can be taken to be a round $S^3$ and with respect to the left-invariant modes on $S^3$ the solution was found to be stable in \cite{Blaback:2011nz}.

\subsection{Supersymmetric $\mathrm{AdS}_7$ vacua from localized D6-branes}

\subsubsection*{System of differential equations}
The pure spinor approach allows one to rewrite the system of supersymmetry equations of type IIA supergravity on the background $\ads$, plus the Bianchi identities for the fluxes, as a system of differential equations involving two differential forms $\psi^1$ and $\psi^2$ (associated with the internal geometry), and the fluxes. According to the Ansatz
\begin{equation} \label{eq:ads7m3}
\d s_{10}^2 = \e^{2A} \d s_{\mathrm{AdS}_7}^2 + \d s^2_{M_3}\, ,
\end{equation} 
the supersymmetry parameters $\epsilon_1$, $\epsilon_2$ (two ten-dimensional Majorana-Weyl spinors with opposite chirality) are decomposed into an external spinor times an internal one: $\epsilon_i = (\zeta \otimes \chi_i \pm \zeta^c \otimes \chi_i^c ) \otimes v_{\pm}$, $i=1,2$, where the superscript $c$ denotes Majorana conjugation and the last factor $v_{\pm}$ is introduced to give $\epsilon_i$ the correct chirality in ten dimensions. The polyforms $\psi^{1,2}$ are defined by the internal spinors $\chi_{1,2}$ via the Clifford map:\footnote{The map allows to identify forms with bispinors: $dx^{m_1} \wedge \ldots \wedge dx^{m_p} \mapsto \gamma_{\alpha\beta}^{m_1 \ldots m_p}$. A slash $\cancel{\phantom{j}}$ over a form denotes its image under the Clifford map, i.e. the associated bispinor.}
\begin{equation} \label{eq:psi12}
\cancel{\psi^1} = \chi_1 \otimes \chi_2^\dagger\, ,\quad \quad \cancel{\psi^2} = \chi_1 \otimes {\chi_2^c}^\dagger\,.
\end{equation}
Together with the total RR flux $F=F_0 + F_2$ allowed by the background, they satisfy the equations
\begin{subequations}\label{eq:sys73}
\begin{align}
	&\d_H {\rm Im} (\e^{3A-\phi} \, \psi^1_+) = -2 \e^{2A-\phi} \, {\rm Re} \psi^1_-\ , 
	\label{eq:73I1}\\
	&\d_H {\rm Re} (\e^{5A-\phi} \, \psi^1_+) = 4 \e^{4A-\phi} \, {\rm Im} \psi^1_-\ ,
	\label{eq:73R1}\\
	&\d_H (\e^{5A-\phi} \,\psi^2_+) = -4i \e^{4A-\phi} \, \psi^2_-\ ,
	\label{eq:732}\\
	& \frac{1}{8} \e^{\phi} \star_3 \lambda F = \d A \wedge {\rm Im} \psi^1_+ + \e^{-A} \,{\rm Re} \psi^1_-\ , \label{eq:73f}\\ 
	&\d A \wedge {\rm Re} \psi^1_- = 0 \label{eq:73dAR}\ , \\
	&(\psi^{1,2}_+,\overline{\psi^{1,2}_-})=-\frac i2  \label{eq:73norm}\ .
\end{align}
\end{subequations} 
Here, $\phi$ is the dilaton and $A$ is the warping factor appearing in \eqref{eq:ads7m3}; $H$ is the NSNS flux, $\d_H = \d - H \wedge$ is the twisted exterior derivative and $\lambda$ acts on a $p$-form as $\lambda \alpha_p = (-)^{\lfloor \frac{p}{2} \rfloor} \alpha_p$. Finally, the subscript $\pm$ on $\psi^{1,2}$ indicates the even (odd) part of the polyform and $\left(\,,\right)$ is the usual Chevalley-Mukai pairing between forms (in particular \eqref{eq:73norm} fixes the norms of the internal spinors to one). To seek for genuine vacuum solutions, every physical field should depend only on $M_3$. 

The system given above is equivalent to $\mathcal{N}=1$ supersymmetry on $\ads$; any of its solutions is by construction a supersymmetric $\mathrm{AdS}_7$ vacuum.\footnote{It can be shown that solutions to \eqref{eq:sys73} are a subclass of solutions of the form $\mathrm{Mink}_6 \times M_4$ (by considering AdS as a warped product of Mink by a line). The system of type II supersymmetry equations on $\mathrm{Mink}_6 \times M_4$ in terms of pure spinors first appeared in \cite{Lust:2010by}.}

\subsubsection*{Parametrization of $\psi^{1,2}$}
In order to solve this system one proceeds by parametrizing the polyforms $\psi^{1,2}$ defined in \eqref{eq:psi12}. The most general parametrization of these forms is obtained (following the lines of \cite{Halmagyi:2007ft}) by noticing that the two internal spinors $\chi_1$ and $\chi_2$ define an identity$\times$identity structure on $T_{M_3} \oplus T^*_{M_3}$, since a norm-1 spinor $\chi$ in three dimensions is able to define a Dreibein $\left\lbrace e_a \right\rbrace_{a=1}^3$ (i.e. an identity structure). The spinors $\chi_{1,2}$ are then expanded on the basis $\left\lbrace \chi, \chi^c \right\rbrace$, the coefficients of this expansion being trigonometric functions of some angles $\theta_1$, $\theta_2$ and $\theta_3$, and the expansion is plugged into \eqref{eq:psi12}, in order to parametrize $\psi^{1,2}$.

Employing this parametrization, the one-form parts of \eqref{eq:73I1}, \eqref{eq:73R1}, \eqref{eq:732} directly {\it determine} the Dreibein on $M_3$ (its components turn out to be combinations of derivatives of the angles), that is they {\it give} its metric $\d s^2_{M_3} = e_a e_a$, \eqref{eq:73f} gives the total RR flux, while the three-form part of the system determines $H$.\footnote{The equation of motion for $F_2$ turns out to be automatically satisfied, as the equation of motion for $H$, whereas the Bianchi identity for $F_2$ is a consequence of the explicit expressions of all fluxes, as determined by the system \eqref{eq:sys73}.} One is left with two genuine ODEs: one is the condition that $F_0$ should be piecewise constant (which is the content of its Bianchi identity) and the other one reads
\begin{equation}\label{eq:413}
x\,\d x = (1+ x^2) \d \phi - (5+x^2) \d A\, ,
\end{equation}
with $x \equiv \cos(\theta_1) \sin(\theta_2)$. Finally, the two-form part of the system imposes $\phi$ to be functionally dependent on $A$ (i.e. $\d A \wedge \d \phi=0$); hence $x$ depends on $A$ too, as imposed by \eqref{eq:413}. 

In particular, the metric determined by the system has the form of a {\it fibration of a round $S^2$ over an interval parametrized by $A$}. However neither $A$ nor the other scalar parameters (the angles) were a priori intended as coordinates on the internal manifold: nevertheless, since the analysis of the system\footnote{That is, the explicit expressions of the metric and of the RR and NSNS fluxes obtained from the system.} has been so far only local, the angles can be promoted to coordinates on $M_3$, while it is wiser to introduce a new coordinate $r$ (defined by $\d r = 4\e^A \frac{\sqrt{1-x^2}}{4x-F_0\, \e^{A+\phi}}\d A$), to parametrize the base of the fibration. 

In these new coordinates the metric reads
\begin{equation}
\d s^2_{M_3} = \d r^2 + \frac1{16}\e^{2A}\,(1-x^2)\d s^2_{S^2} \ ,\label{eq:met-r}
\end{equation}
and the system of residual ODEs is:
\begin{subequations}\label{eq:oder}
\begin{align}
	&\partial_r \phi = \frac{\e^{-A}}{4\sqrt{1-x^2}} (12 x + (2x^2-5)F_0 \,\e^{A+\phi}) \ ,\\
	&\partial_r x = -\frac{1}{2} \e^{-A}\sqrt{1-x^2} (4+x F_0 \, \e^{A+\phi}) \ ,\\
	&\partial_r A = \frac{4\e^{-A}}{\sqrt{1-x^2}} (4x - F_0\,  \e^{A+\phi})\ .
\end{align}
\end{subequations}
Now $r$ has become a coordinate on the base and $A$, $x$, $\phi$ have become functions of $r$. Moreover, to make $M_3$ compact, the $S^2$ fiber is demanded to shrink at two distinct values of $r$ (this is accomplished if $x$ goes to $\pm 1$ at the extrema of the base interval, see \eqref{eq:met-r}): thus, $M_3$ is topologically an $S^3$ and these two values are interpreted as its north, $r_{\mathrm{N}}$, and south, $r_{\mathrm{S}}$, poles.\footnote{The other possible compact $M_3 = S^1 \times S^2$ (topologically), which is obtained compactifying the base interval, is excluded since incompatible with the system \eqref{eq:oder}.} In turn, the compact topology of the internal space imposes boundary conditions on the system, that must be satisfied by the fields.

A way of understanding the appearance of this $S^2$ is by considering the $\mathrm{Sp}(1) \cong \mathrm{SU}(2)$ R-symmetry group of a six-dimensional $(1,0)$ CFT, dual to any possible solution of the system; furthermore, its presence elucidates why the even- and odd-form parts of the bispinors $\psi^{1,2}$ can be organized as singlets and triplets of $\mathrm{SU}(2)$ \cite[Sec.~4.5]{Apruzzi:2013yva}.

\subsubsection*{Massive solutions with localized D6-branes}
Assuming $F_0 \neq 0$ everywhere on $M_3$, it is possible to construct a compact solution to \eqref{eq:oder} with nonzero D6 charge. Such a fully backreacted solution contains a stack of spacetime-filling D6-branes localized at the south pole (the stack being there calibrated). Around this pole the metric is singular since it has the behavior one expects near a D6, and the $H$ flux is divergent; nontrivially, the solution exhibits peculiar global properties: the north pole is a regular point for all fields, but one can also substitute it by inserting an anti-D6 stack or an O6-plane still obtaining a globally well-defined solution (in these cases too, $H$ diverges near the sources).

\subsection{Smearing breaks supersymmetry}
We now prove that the system \eqref{eq:sys73} does not allow for any solution with smeared D6 charge. Smearing the system practically means that we will enforce the following conditions:
\begin{equation} \label{eq:smear}
\phi = \mathrm{const} \equiv \phi_0,\quad A=\mathrm{const},\quad F_2 = 0\ .
\end{equation}
Since for constant $A$ the warping factor $e^{2A}$ appearing in \eqref{eq:ads7m3} can be reabsorbed in the $\mathrm{AdS}_7$ metric, we will set $A=0$.  Thus, the ten-dimensional metric takes the form of a direct product: 
\begin{equation}\label{eq:smemet}
\d s_{10}^2 = \d s^2_{\mathrm{AdS}_7} + \d s^2_{M_3}\ .
\end{equation}
No condition is imposed on $H$ and $F_0$ so that, a priori, they are not identically zero. Under these assumptions the system \eqref{eq:sys73} simplifies while \eqref{eq:73norm} holds unchanged.

From \eqref{eq:413} and imposing \eqref{eq:smear}, we find
\begin{equation} \label{eq:reldif}
\tan(\theta_1)\,  \d \theta_1 = \cot(\theta_2)\, \d \theta_2\ .
\end{equation}
This is a nonlinear relation between two of the differentials (derivatives of the angles) which induces a relation between two of the components of the Dreibein. Therefore, if we assume that $\left\lbrace e_a\right\rbrace$ is a well-defined Dreibein on $M_3$, the one-form equations cannot be solved together. We have thus shown that smearing breaks supersymmetry, since the smeared system does not allow for any supersymmetric solution.

Nevertheless, it is possible to define the smeared limit of the massive solution with D6-branes of \cite{Apruzzi:2013yva} as a bona fide solution to the ten-dimensional equations of motion where delta sources are replaced by constants \cite{Blaback:2011nz, Blaback:2011pn}; it just breaks supersymmetry. 

\subsection{Non-supersymmetric $\mathrm{AdS}_7$ vacua and vacuum stability}

Before we move towards the gauged supergravity analysis we want to address a few issues related to the existence of possible non-supersymmetric extensions of these $\mathrm{AdS}_7$ vacua and spend a few words on the stability of the latter. Our aim is to connect \cite{Apruzzi:2013yva} to \cite{Blaback:2011pn, Blaback:2012nf, Bena:2012tx}.

Concerning non-supersymmetric extensions, one is required to solve the general second order differential equations for an Ansatz that has rotational symmetry \cite{Blaback:2011pn}. In Einstein frame the Ansatz is given by:
\begin{align}
& \d s^2_{10} = \e^{2A(\theta)} \,\d s_{\mathrm{AdS}_7}^2 + \e^{2B(\theta)}\left(\d \theta^2 + \sin^2(\theta)
\,\d \Omega^2\right)\ ,\\
& H=\lambda F_0\, \e^{\tfrac{7}{4}\phi}\star_3 1\ , \label{eq:Hflux}\\
& F_2 = \e^{-\tfrac{3}{2}\phi-7A}\star_3\d\alpha\ ,
\end{align}
where $\phi, \lambda$  and $\alpha$ are now functions depending on $\theta$, $\star_3$ contains the conformal factor and we take $F_0$ to be constant.  The equation of motion for $H$ enables one to eliminate $\alpha$ in terms of $\lambda$,
\begin{equation}\label{alphaversuslambda}
\alpha=\e^{\tfrac{3}{4}\phi +7A}\, \lambda \ ,
\end{equation}
where we have set the integration constant to zero by shifting $\alpha$. Then the problem is reduced to finding a set of four unknown functions $A, B, \phi, \lambda$ depending on $\theta$ and obeying coupled second order differential equations. Around $\theta =0$ the general solution is given by \cite{Blaback:2011pn}:
\begin{align}
& \e^{-A} = \theta^{-\frac{1}{16}}\Bigl(a_0 + a_1 \theta + \ldots \Bigr)\ ,\\
& \e^{-2B}= \theta^{\frac{7}{8}}\Bigl(b_0 + b_1 \theta + \ldots \Bigr)\ ,\\
& \e^{-\frac{1}{4}\phi}= \theta^{-\frac{3}{16}}\Bigl(f_0 + f_1 \theta + \ldots \Bigr)\ ,\\
& \lambda = \theta^{-1}\Bigl(\lambda_0 + \lambda_1 \theta + \ldots \Bigr)\ .
\end{align}
To understand what the general integration parameters are one investigates which of the Taylor expansion coefficients can be chosen freely. It turns out that there are five constants and the rest can be determined in terms of these five \cite{Blaback:2011pn}:
\begin{equation}\label{five}
a_0,\ b_0,\ f_0,\ \lambda_0,\ \lambda_1\ .
\end{equation}
The reason can easily be understood. The ten-dimensional equations of motion can be interpreted as four coupled differential equations plus a Hamiltonian constraint. This would give seven integration constants. However $A(0)$ and $B(0)$ can be understood as rescaling the $\mathrm{AdS}_7$ and $S^3$ radii such that we take them equal to zero. By fixing the D6 charge at the origin $\theta=0$ we enforce one algebraic condition among the constants in (\ref{five}), such that one is left with four independent integration constants.  What was not done in \cite{Blaback:2011pn} is to check which of these local solutions can be extended consistently all the way down to the south pole. The way this should proceed is via the shooting method. One constructs the solutions near the north and south pole and evolves them towards the equator where they have to connect smoothly. We expect that this introduces four extra constraints on the above integration constants, one for each degree of freedom ($A$, $B$, $\alpha$, $\phi$). This then fixes the solution \emph{uniquely} and implies that all the solutions to the second order equations must be supersymmetric, when one demands them to be globally well-defined.\footnote{This means that the BPS conditions of \cite{Apruzzi:2013yva} are required for a globally well-defined solution. This fixes for instance $\lambda_0 = \frac{a_0 f_0^5}{F_0}$, which can also be seen by comparing with the expression $H=-(6\e^{-A}+xF_0\,\e^{\phi})\mathrm{vol}_3$ given in \cite{Apruzzi:2013yva}.} This shows that the solutions have no moduli, as already noticed in \cite{Apruzzi:2013yva}, and that the solution is completely fixed by discrete topological data: the Romans mass $F_0$, the total flux integer $h=\int H$ and the way the D6 charges are distributed over north and south pole of the $S^3$. The total D6 charge $Q_6$ is determined by the RR tadpole condition:
\begin{equation}
Q_6 = Q_{\text{south}} + Q_{\text{north}} = h F_0\ . 
\end{equation} 

Finally we address the issue of stability of the supersymmetric $\mathrm{AdS}_7$ solutions. Despite supersymmetry, one needs to worry about stability because of the $H$ singularity. As long as this singularity is not resolved the solution is not physical and one possible interpretation is that such a solution is simply not existent and one should really have a time-dependent solution that describes flux decaying against the D6 charges \cite{Blaback:2012nf}.\footnote{Supersymmetry is not a guarantee for existence and stability when singularities are present. Well known examples are multi-centered BPS black holes \cite{Denef:2000nb}. When all centers are brought together one finds a BPS spherical solution with naked singularity. This solution is not physical and wants to evolve in time, separating the centers until they relax to their equilibrium position of the multi-centered solution.} The essential ingredient that decides on the fate of the solution is the resolution of the singularity. If the D6-branes can be replaced with spherical D8 shells that carry the same D6 charge then the singularity disappears \cite{Bena:2012tx}. This has been successful for those $\mathrm{AdS}_7$ solutions which do not carry net D6 charge ($Q_{\text{south}} =- Q_{\text{north}}$) \cite{Apruzzi:2013yva}. For the other solutions, a probe analysis in the noncompact limit has revealed that the polarization is not occuring \cite{Bena:2012tx}. However this does not exclude that in the compact case, at some far enough distance from the pole, the formation of spherical D8-branes could occur. In fact, from a holographic point of view one is tempted to conclude that the class of $\mathrm{AdS}_7$ solutions with nonzero Romans mass could be dual to $(1,0)$ CFT's in $D=6$ \cite{Hanany:1997gh}. The existence of such CFT's would rule out the possibility of having an unstable vacuum and hence one expects D8-branes to polarize and resolve the singularity on the gravity side.  However it is essential to understand that this does not relate to the fate of supersymmetry-breaking anti-branes in warped throats. It is only the noncompact version of these compact $\mathrm{AdS}_7$ solutions, for which the worldvolume is Minkowski instead of AdS, that the D6 sources can be interpreted in terms of supersymmetry-breaking branes, that decay into noncompact ten-dimensional Minkowski spacetime \cite{Blaback:2012nf}.

\section{Gauged supergravity in $D=7$ and IIA compactifications}\label{gauged}

In this section we attempt to interpret a class of massive type IIA compactifications with smeared six-branes as gauged supergravities in $D=7$, possibly up to some explicit supersymmetry-breaking effects in the corresponding scalar potential. At first glance, since we are including half-BPS objects, one might think that half-maximal supergravity theories are the correct framework to analyze this problem.

However, since the $\mathbb{Z}_{2}$ truncation worked out in \cite{Dibitetto:2012rk} relating the maximal theory to the half-maximal one can be interpreted as an O6 involution, only orientifold-allowed fluxes can be described by using the embedding tensor of the half-maximal theory. In particular, within such a framework, we will only be able to describe a compactification carrying nonzero Romans mass and NSNS three-form flux, but no ``metric flux''. In order to include this, we will need the full embedding tensor of the maximal theory.

The reduction Ansatz, in string frame, that produces the result which we will be comparing our supergravity potentials with, reads
\be
\d s^2_{10} \, = \, \tau^{-2} \d s_7^2 \, + \, \rho \, M_{ij}e^{i} \otimes e^{j} \ , 
\ee
where $\rho$ and $\tau$ are suitable combinations of the internal volume and ten-dimensional dilaton guaranteeing that the seven-dimensional Lagrangian is in the Einstein frame, whereas $M_{ij}$ parametrizes the $\textrm{SL}(3)/\textrm{SO}(3)$ coset, where $i, \, j \, = \, 1, \, 2, \, 3$ denote $\textrm{SL}(3)$ fundamental indices. The $e^i$'s are Maurer--Cartan one-forms and the structure constants of their algebra are denoted by $\omega$.

By performing the reduction, one finds the following scaling properties for the different fluxes
\be
\begin{array}{lclclc}
V_{F_{0}} \, \sim \, f_{0}^{2} \, \rho^{3/2} \, \tau^{-7} & , & V_{H} \, \sim \, h^{2} \, \rho^{-3} \, \tau^{-2} & , & V_{\omega} \, \sim \, \omega^{2} \, \rho^{-1} \, \tau^{-2} & ,
\end{array}
\ee
which imply that the above fluxes naturally transform in the following irreps of $\mathbb{R}^{+}_{\rho} \, \times \, \mathbb{R}^{+}_{\tau} \, \times \, \textrm{SL}(3)$:
\be
\label{Flux_irreps}
\hspace{-3mm}
\begin{array}{lclclc}
F_{0} \, =  \, f_{0} \, \in \, \textbf{1}_{(-\frac{3}{4};\,+\frac{7}{2})} & , & H_{ijk} \, =  \, h \, \epsilon_{ijk} \, \in \, \textbf{1}_{(+\frac{3}{2};\,+1)} & , & 
{\omega_{ij}}^{k} \, =  \, \epsilon_{ijl} \, q^{lk} \, \in \, \textbf{6}^{\prime}_{(+\frac{1}{2};\,+1)} & ,
\end{array}
\ee
where $q^{(ij)} \, = \, q^{ij}$. 

\subsection{Half-maximal gauged supergravity}

Seven-dimensional half-maximal supergravity enjoys $\mathbb{R}^{+} \, \times \, \textrm{SL}(4)$ global symmetry. Note that $\mathrm{SL}(4) \cong \mathrm{SO}(3,3)$ and hence we are here only considering the
theory coupled to three vector multiplets that are included in the closed-string sector. The extra $N$ arbitrary vector multiplets which contain open-string degrees of freedom (see appendix \ref{IIACOMP}) do not play any essential role in this analysis and hence including them would not change the corresponding result. 

The consistent deformations of this theory transform as
\be
\label{ET_half}
\begin{array}{cccccc}
\Theta & \in & \underbrace{\textbf{1}_{(-4)}}_{p=3} & \oplus & \underbrace{\textbf{6}_{(+1)}\,\oplus\,\textbf{10}_{(+1)}\,\oplus\,\textbf{10}^{\prime}_{(+1)}}_{p=1} & ,
\end{array}
\ee
where the subscripts on the different $\textrm{SL}(4)$ irreps denote $\mathbb{R}^{+}$ charges. This is in agreement with what predicted in \cite{Bergshoeff:2007vb} by using the Kac-Moody approach, where it is also shown that the $\textbf{1}_{(-4)}$ corresponds to a ``$p=3$-type'' deformation, i.e. a St\"uckelberg-like massive deformation for the three-form field, thus not associated with any gauging. The other irreps instead correspond to gaugings; the $\textbf{6}_{(+1)}$ gauges the $\mathbb{R}^{+}$ factor and some subgroup of $\textrm{SL}(4)$, whereas gaugings in the $\textbf{10}_{(+1)}\,\oplus\,\textbf{10}^{\prime}_{(+1)}$ are purely within $\textrm{SL}(4)$. 

In what follows we will show how only the Romans mass and $H$ flux coming from compactifications of massive IIA supergravity with six-branes sit inside the  deformations $\Theta$ introduced in \eqref{ET_half}. To this end, we will restrict ourselves to the relevant case\footnote{\label{foot_xi}The only known ten-dimensional construction giving rise to $\mathbb{R}^{+}$ gaugings parametrized by an embedding tensor in the $\textbf{6}_{(+1)}$ is introducing some dilaton flux $H_{i} \, \equiv \, \partial_{i}\phi$. This would turn on at most half of its components, but it goes beyond our present scope.} of purely $\textrm{SL}(4)$ gaugings combined with a massive deformation. 

Let us denote by $\theta \, \in \, \textbf{1}_{(-4)}$ the mass parameter and by $Q_{(mn)} \, \in \, \textbf{10}_{(+1)}$ and $\tilde{Q}^{(mn)} \, \in \, \textbf{10}^{\prime}_{(+1)}$ the embedding tensor components, where $m, n$ are fundamental $\textrm{SL}(4)$ indices. In this case, the non-vanishing Quadratic Constraints (QC) needed for the closure of the gauge algebra reduce to \cite{Dibitetto:2012rk}
\be
\label{QC_half}
\begin{array}{cccc}
\tilde{Q}^{mp} \, Q_{pn} \, - \, \dfrac{1}{4} \, \left(\tilde{Q}^{pq} \, Q_{pq}\right) \, \delta_{n}^{m} & = & 0 & , 
\end{array}
\ee
transforming in the $\textbf{15}_{(+2)}$ of $\textrm{SL}(4)$. 

In terms of the scalars of the theory, which span 
\be
\begin{array}{cccc}
\underbrace{\mathbb{R}^{+}}_{\Sigma} & \times & \underbrace{\frac{\textrm{SL}(4)}{\textrm{SO}(4)}}_{\mathcal{M}_{mn}} & , 
\end{array}
\ee
the scalar potential induced by the above deformation parameters can be written as\footnote{Please note that we have chosen the normalization of the mass parameter $\theta$ such that the gauge coupling $g$ factorizes the whole potential given in \eqref{V_Half_Max}.}
\be
\label{V_Half_Max}
\begin{array}{ccrlc}
V & = & \dfrac{g^{2}}{64} & \bigg[\theta^{2} \, \Sigma^{8} \, + \, \dfrac{1}{4} \, Q_{mn} \, Q_{pq} \, \Sigma^{-2} \, \left(2 \, \mathcal{M}^{mp} \, \mathcal{M}^{nq} \, - \, \mathcal{M}^{mn} \, \mathcal{M}^{pq} \right) 
 & + \\[3mm]
& & + & \dfrac{1}{4} \, \tilde{Q}^{mn} \, \tilde{Q}^{pq} \, \Sigma^{-2} \, \left(2 \, \mathcal{M}_{mp} \, \mathcal{M}_{nq} \, - \, \mathcal{M}_{mn} \, \mathcal{M}_{pq} \right)
 & + \\[3mm]
& & - & \theta \, \left(\tilde{Q}^{mn} \, \Sigma^{3} \, \mathcal{M}_{mn} \, + \,  Q_{mn} \, \Sigma^{3} \, \mathcal{M}^{mn}\right) \, + \, Q_{mn} \, \tilde{Q}^{mn} \, \Sigma^{-2}\bigg] & ,
\end{array}
\ee
where $\mathcal{M}^{mn}$ denotes the inverse of $\mathcal{M}_{mn}$. The first step to obtain such an expression for the scalar potential $V$ is $\mathbb{Z}_{2}$ truncating the maximal theory \cite{Samtleben:2005bp} as described in \cite{Dibitetto:2012rk}. This gives rise to a particularly constrained half-maximal theory where the following extra QC are satisfied
\be
\label{QC_extra}
\begin{array}{ccccccccccc}
\tilde{Q}^{pq} \, Q_{pq} & = & 0 & & \textrm{and} & & & \theta \, \tilde{Q}_{mn} & = & 0 & .
\end{array}
\ee
Subsequently one can observe that the general scalar potential of the half-maximal theory will contain the two above terms with some coefficients. Finally, one can fix those coefficients by performing
a $\mathbb{T}^{2}$ reduction down to $D=5$ and comparing the result with the corresponding terms in the scalar potential of \cite{Schon:2006kz}.

The relation between embedding tensor components and fluxes reads:\footnote{See appendix~\ref{app:dictionary} for some details concerning the derivation.}
\be
\label{dictionary_fluxes_half}
\begin{array}{lclc}
\theta \, =  \, - \dfrac{1}{\sqrt{2}} \, h & , & \tilde{Q}^{00} \, =  \, \sqrt{2} \, f_{0} & .
\end{array}
\ee

Concerning the scalar sector, the $\textrm{SL}(3)$ sector parametrized by $M_{ij}$, which is naturally obtained by dimensional reduction, is embedded inside $\mathcal{M}_{mn}$ in the
following way
\be
\label{SL4_scalars}
\mathcal{M}_{mn} \, = \, \left(
\begin{array}{c | c}
\Phi^{3} & \\
\hline
 & \\[-2mm]
 & \Phi^{-1} \, M_{ij}
\end{array}\right) \ .
\ee

Now, by inserting the parametrization of the $\textrm{SL}(4)$ scalars given in \eqref{SL4_scalars} and the dictionary \eqref{dictionary_fluxes_half} in the expression of the scalar potential \eqref{V_Half_Max}, one finds
\be
V \, = \, \frac{g^{2}}{128} \, \left(\frac{h \, \Sigma^{5} \, + \, f_{0} \, \Phi^{3}}{\Sigma}\right)^{2} \ .
\ee
This coincides with the expression obtained in \cite{Blaback:2011nz} adapted to the case with no metric flux,\footnote{Please note that we have adopted different conventions w.r.t. \cite{Blaback:2011nz}, where $V$ is obtained from a reduction in the string frame, thus directly being a function of the ten-dimensional dilaton $\phi$ and the volume modulus $v$.}
\be
\label{V_Fluxes_half}
V \, = \, \frac{\left(h \, \tau^{5/2} \, + \, f_{0} \, \rho^{9/4}\right)^{2}}{2 \, \rho^{3} \, \tau^{7}} \ ,
\ee
by choosing $g \, = \, 8$, $h \, f_{0}$ equal to the D6 (or anti-D6, depending on its sign) tension and upon using the following mapping between the $\mathbb{R}^{+}$ scalars
\be
\label{dictionary_scalars}
\begin{array}{lclccclclc}
\Sigma & \equiv & \rho^{-3/8} \, \tau^{-1/4} & , & & \Phi & \equiv & \rho^{1/8} \, \tau^{-5/4} & 
\end{array}
\ee
(which can be derived as a consequence of \eqref{dictionary_weights}). When the D6 tension is negative, as is the case for O6-planes, then there is a stable Minkowski vacuum for those values of the fields such that $h \, \tau^{5/2} \, + \, f_{0} \, \rho^{9/4}=0$. At this Minkowski point a certain combination of $\rho$ and $\tau$ remains massless. This seven-dimensional gauged supergravity with a no-scale structure is discussed in detail in appendix \ref{IIACOMP}.

This Minkowski solution, as a solution to seven-dimensional gauged supergravity, solves the ten-dimensional equations of motion in the smeared O6 case. But the warped version, with fully localized O6-planes, is known as well \cite{Blaback:2010sj}.\footnote{It would be useful to use this vacuum solution as an explicit background to investigate some of the issues raised in \cite{McOrist:2012yc} and \cite{Saracco:2012wc}.} Even more, it is possible to map the BPS domain wall flows in that gauged supergravity to ten-dimensional solutions with localized O6-planes \cite{Blaback:2012mu, toappear}. The analysis in \cite{Blaback:2012mu, toappear} shows a perfect match between the  conditions in the seven-dimensional gauged supergravity from smeared O6-planes and the ten-dimensional supersymmetry conditions with localized O6-planes. 
This matching is generally to be expected and this is why we consider it worthy to emphasize that the $\mathrm{AdS}_7$ solutions in massive IIA display the opposite behaviour.

\subsection{Maximal gauged supergravity}

Now let us move to the maximal theory which will allow us to include metric flux in our discussion.
Maximal supergravity in $D=7$ enjoys $\textrm{SL}(5)$ global symmetry. The consistent deformations of this theory (all corresponding to gaugings) transform as \cite{Samtleben:2005bp}
\be
\label{ET}
\begin{array}{cccccc}
\Theta & \in & \underbrace{\textbf{15}}_{Y_{MN}} & \oplus & \underbrace{\textbf{40}^{\prime}}_{Z^{MN,P}} & ,
\end{array}
\ee
where the Linear Constraint (LC) implies $Y_{(MN)} \, = \, Y_{MN}$ and $Z^{[MN],P} \, = \, Z^{MN,P}$ with $Z^{[MN,P]} \, = \, 0$, where $M, \, N, \, P$ denote fundamental $\textrm{SL}(5)$ indices.
The following Quadratic Constraints (QC) are needed for the closure of the gauge algebra 
\be
\label{QC_max}
\begin{array}{cccc}
Y_{MQ} \, Z^{QN,P} \, + \, 2 \, \epsilon_{MRSTU} \, Z^{RS,N} \, Z^{TU,P} & = & 0 & , 
\end{array}
\ee
transforming in the $\textbf{5}^{\prime} \, \oplus \, \textbf{45}^{\prime} \, \oplus \, \textbf{70}^{\prime}$ of $\textrm{SL}(5)$. 

The scalars of the theory describe fourteen propagating degrees of freedom and are parametrized by an element $\mathcal{M}_{MN}$ of the coset $\textrm{SL}(5)/\textrm{SO}(5)$.
The embedding tensor deformations introduced in \eqref{ET} induce the following scalar potential
\be
\label{V_Max}
\begin{array}{ccrlc}
V & = & \dfrac{g^{2}}{64} & \bigg[Y_{MN} \, Y_{PQ} \, \left(2 \, \mathcal{M}^{MQ} \, \mathcal{M}^{NP} \, - \, \mathcal{M}^{MN} \, \mathcal{M}^{PQ} \right) & + \\[3mm]
& & + &  64 \, Z^{MN,P} \, Z^{QR,S} \, \mathcal{M}_{MQ} \, \left(\mathcal{M}_{NR} \, \mathcal{M}_{PS} \, - \, \mathcal{M}_{NP} \, \mathcal{M}_{RS} \right)\bigg] & .
\end{array}
\ee

In what follows we will construct the dictionary between the above deformation parameters $\Theta$ and fluxes in compactifications of massive IIA supergravity with D6-branes.  
To this end, we will restrict ourselves to those embedding tensor components which have some ten-dimensional origin in this duality frame.

The explicit form of the dictionary embedding tensor/fluxes reads:
\be
\label{dictionary_fluxes_max}
\begin{array}{lclclc}
Y_{++} \, =  \, 4\sqrt{2} \, h & , & Z^{-+,-} \, =  \, - Z^{+-,-}\, =  \, \dfrac{1}{\sqrt{2}} \, f_{0} & , & Z^{i-,j} \, =  \, - Z^{-i,j} \, =  \dfrac{1}{\sqrt{2}} \, q^{ij} & .
\end{array}
\ee

For what concerns the scalar sector, $\Sigma$, $\Phi$ and $M_{ij}$ are embedded in the following way inside the element $\mathcal{M}_{MN}$ of the $\textrm{SL}(5)/\textrm{SO}(5)$ coset: 
\be
\label{SL5_scalars}
\mathcal{M}_{MN} \, = \, \left(
\begin{array}{c | c | c}
\Sigma^{-4} &  & \\
\hline
 & \\[-2mm]
& \Sigma \, \Phi^{3} & \\
\hline
 & \\[-2mm]
 &  & \Sigma \, \Phi^{-1} \, M_{ij}
\end{array}\right) \ .
\ee

Now, by inserting the parametrization of the $\textrm{SL}(5)$ scalars given in \eqref{SL5_scalars} and the dictionary \eqref{dictionary_fluxes_max} in the expression of the scalar potential 
\eqref{V_Max}, one finds:
\be
V \, = \, \frac{g^{2}}{2} \, \left[h^{2} \, \Sigma^{5} \, + \, f_{0}^{2} \, \Sigma^{-2} \, \Phi^{6} \, + \, \Sigma^{3} \, \Phi \, \left(2\, \textrm{Tr}(q\,M\,q\,M) \, 
- \, \textrm{Tr}(q\,M)^{2}\right)\right] \ .
\ee 
By making use of the dictionary \eqref{dictionary_scalars} for the $\mathbb{R}^{+}$ scalars to compare the above expression with\footnote{The following expression was obtained in \cite{Blaback:2011nz} by means of a reduction of massive IIA supergravity with smeared six-branes.}
\be
\label{V_Fluxes_max}
V \, = \, \frac{\left(h \, \tau^{5/2} \, + \, f_{0} \, \rho^{9/4}\right)^{2}}{2 \, \rho^{3} \, \tau^{7}} \, + \, \rho^{-1} \, \tau^{-2} \, \left(\textrm{Tr}(q\,M\,q\,M) \, 
- \, \frac{1}{2} \, \textrm{Tr}(q\,M)^{2}\right)\ ,
\ee
one finds that they only coincide when the $h \, f_{0}$
term corresponding to the tadpole generated by the smeared sources is absent.

\subsection*{Summarizing}

The scalar potential given in \eqref{V_Fluxes_max} coming from reductions of massive IIA supergravity with smeared D6 charge can be written in two different ways:
\be
\begin{array}{lclclc}
V_{(\textrm{IIA})} & = & V_{(\textrm{half-max.})} \, + \, \omega^{2} \, \rho^{-1} \, \tau^{-2} & = & V_{(\textrm{max.})} \, + \, T_{6} \, \rho^{-3/4} \, \tau^{-9/2} & ,
\end{array}
\ee
where $\omega^{2} \, \propto \, \left(\textrm{Tr}(q\,M\,q\,M) \, - \, \frac{1}{2} \, \textrm{Tr}(q\,M)^{2}\right)$ and $T_{6} \, \propto \, h \, f_{0}$.

Both maximal and half-maximal gauged supergravities miss out a term in the scalar potential $V_{(\textrm{IIA})}$. Since these are the only existing consistent supergravity theories in $D=7$, we conclude that the scalar potentials coming from this class of compactifications do not admit any gauged supergravity description in general.

\section{Discussion} \label{Discussion}
The main message of this paper is twofold. First we observe that there exists an $\mathrm{AdS}_7$ flux vacuum with sixteen supercharges that has no gauged supergravity description in seven dimensions. Second, this supersymmetric vacuum is such that supersymmetry is broken when the spacetime-filling branes, required for the existence of the vacuum, are smeared over the internal manifold. These two observations are related since gauged supergravity descriptions are typically obtained from smearing branes and orientifold planes. However this would still allow the possibility that the flux vacuum be a $\mathcal{N}=0$ solution in a seven-dimensional gauged supergravity. We have verified that this cannot be the case.

The practical obstacle for the smeared $\mathrm{AdS}_7 \times S^3$ solution to be part of a seven-dimensional gauged supergravity is the $S^3$ geometry. Although $S^3$ is a group manifold and the smeared $\mathrm{AdS}_7$ solution has only left-invariant modes excited, it turns out that calibrated smeared D6-branes are only allowed in flat internal geometries. We have shown this using the embedding tensor formalism.  The essential mechanism behind this is the observation that compactifications with spacetime-filling sources and extended supersymmetry require the O$p$ involutions \cite{Dibitetto:2011eu, Dibitetto:2012rk}, even when the only sources would be D$p$-branes. In our case O6 involutions would project out the ``metric flux'' of the $S^3$ and only allow $\mathbb{T}^3$ as an internal geometry. 

The simple observations made in this note illustrate that gauged supergravity is a restrictive tool when it comes to classifying flux vacua, even when these vacua preserve many supercharges. 

It is natural to wonder about the existence of the effective field theory description of the low-energy fluctuations around the $\mathrm{AdS}_7$ vacuum, since one would naively expect this to be given by a half-maximal gauged supergravity. The reason this is not the case is the absence of a parametric separation between the $\mathrm{AdS}_7$ curvature radius and the KK scale. This absence implies that there is no lower-dimensional effective field theory. An observer in this spacetime will always see all of the ten spacetime dimensions. This is similar to the standard Freund-Rubin vacua, although they admit a gauged supergravity description. But these gauged supergravities should not be regarded as effective field theories. They are rather consistent truncations of ten-dimensional degrees of freedom that combine into lower-dimensional supergravity multiplets, since FR vacua are always perceived as higher-dimensional to an observer. Therefore we conjecture that our observations cannot occur for compactifications that are genuinly lower-dimensional in the sense of a parametric scale separation between the AdS curvature radius and the KK radius. Such supersymmetric AdS vacua should always be obtainable from lower-dimensional supergravities and smearing should not break supersymmetry. 

\subsection*{Acknowledgements}
We have benefited from useful correspondence with Frederik Denef, Adolfo Guarino, Joe Minahan, Alessandro Tomasiello and Marco Zagermann.  The work of U.D. and G.D. is supported by the Swedish Research Council (VR), and the G\"oran Gustafsson Foundation. The work of M.F.~was partially supported by the ERC Advanced Grant ``SyDuGraM'', by IISN-Belgium (convention 4.4514.08) and by the ``Communaut\'e Fran\c{c}aise de Belgique" through the ARC program. M.F.~ is a Research Fellow of the Belgian FNRS-FRS. The work of T.V.R. is supported by a Pegasus Marie Curie fellowship of the FWO.

\appendix

\section{The dictionary embedding tensor/fluxes}
\label{app:dictionary}

In this appendix we would like to spell out the details of the group-theoretical computation that produces the dictionary between fluxes and embedding tensor components given in \eqref{dictionary_fluxes_half}
 and \eqref{dictionary_fluxes_max}, respectively in the case of half-maximal and maximal gauged supergravity in seven dimensions.

\subsection*{The half-maximal case}

From the branching of the embedding tensor $\mathbb{R}^{+}_{\Sigma} \, \times \, \textrm{SL}(4)$ irreps introduced in \eqref{ET_half}, w.r.t. its 
$\mathbb{R}^{+}_{\Sigma} \, \times \, \mathbb{R}^{+}_{\Phi} \, \times \, \textrm{SL}(3)$ subgroup\footnote{We adopt the following notation $m \rightarrow 0 \oplus i$, and keep no embedding tensor transforming in the $\textbf{6}_{(+1)}$ of $\mathbb{R}^{+}_{\Sigma} \, \times \, \textrm{SL}(4)$ (see footnote~\ref{foot_xi}).}
\be
\begin{array}{lclc}
\textbf{1}_{(-4)} & \longrightarrow & \textbf{1}_{(-4; \,0)} & , \\[2mm]
\textbf{10}_{(+1)} & \longrightarrow & \textbf{1}_{(+1; \,+3)} \, \oplus \, \textbf{3}_{(+1; \,+1)} \, \oplus \, \textbf{6}_{(+1; \,-1)} & , \\[2mm] 
\textbf{10}^{\prime}_{(+1)} & \longrightarrow & \textbf{1}_{(+1; \,-3)} \, \oplus \, \textbf{3}^{\prime}_{(+1; \,-1)} \, \oplus \, \textbf{6}^{\prime}_{(+1; \,+1)} & ,
\end{array} \notag
\ee
one realizes that there is an ambiguity in placing the two $\textrm{SL}(3)$ singlets representing $h$ and $f_{0}$ inside the three deformation parameters.  
This is however solved by comparing with the four-dimensional dictionary \cite{Dibitetto:2011gm} after a $\mathbb{T}^{3}$ reduction. One finds that none of them is sitting inside $Q$, whereas the other
two irreps contain $H$ and $F_{0}$ fluxes as shown in \eqref{dictionary_fluxes_half}.

By subsequently plugging \eqref{dictionary_fluxes_half} into the QC given in \eqref{QC_half}, one finds that no condition needs to be imposed for consistency. This is in agreement with what predicted by 
dimensional reduction.
Moreover, one more interesting check consists in working out the extra QC \eqref{QC_extra} required in order to have an uplift to maximal supergravity in this case. What one finds is 
$h \, f_{0} \, = \, 0$, which is in perfect agreement with the prediction that maximal supersymmetry should not allow for any D6-branes. 

This also allows one to fix the dictionary between the scaling weights w.r.t. the supergravity $\mathbb{R}^{+}$ scalars and the $\rho$ and $\tau$ scalings given in \eqref{Flux_irreps}
\be
\label{dictionary_weights}
\begin{array}{lclccclclc}
q_{\Sigma} & = & -\frac{5}{2} q_{\rho} \, - \, \frac{1}{4} q_{\tau} & , &  & q_{\Phi} & = & \frac{1}{2} q_{\rho} \, - \, \frac{3}{4} q_{\tau} & ,
\end{array}
\ee
which produces the mapping in \eqref{dictionary_scalars}.
According to this dictionary, metric flux ${\omega_{ij}}^{k}$ is expected to transform in the $\textbf{6}^{\prime}_{(-\frac{3}{2};-\frac{1}{2})}$. Since there is no object having the correct scaling in the above decomposition, one can conclude that metric flux cannot be included within the embedding tensor of the half-maximal theory. 

\subsection*{The maximal case}

In order to perform the flux analysis in this case, we need to branch the $\textrm{SL}(5)$ embedding tensor w.r.t. its $\mathbb{R}^{+} \, \times \, \mathbb{R}^{+} \, \times \, \textrm{SL}(3)$ subgroup. One finds\footnote{We adopt the following notation: $M \rightarrow +  \oplus -  \oplus i$. }
\be
\begin{array}{cclc}
\textrm{SL}(5) & \supset & \qquad \mathbb{R}^{+}_{\Sigma} \, \times \, \mathbb{R}^{+}_{\Phi} \, \times \, \textrm{SL}(3) & \\[3mm]
\textbf{15} & \longrightarrow & \quad \, \textbf{1}_{(+1;+3)} \, \oplus \, \textbf{3}_{(+1;+1)} \, \oplus \, \textbf{6}_{(+1;-1)} \, \oplus \,
\underline{\textbf{1}_{(-4;\,0)}} \, \oplus \, \textbf{1}_{(-\frac{3}{2};+\frac{3}{2})}  \, \oplus \, \textbf{3}_{(-\frac{3}{2};-\frac{1}{2})}  & , \\[4mm]

\textbf{40}^{\prime} & \longrightarrow & \left\{\begin{array}{c} \underline{\textbf{1}_{(+1;-3)}} \, \oplus \, \textbf{3}^{\prime}_{(+1;-1)} \, \oplus \, \textbf{6}^{\prime}_{(+1;+1)} \, \oplus \,
\textbf{1}_{(+\frac{7}{2};-\frac{3}{2})} \, \oplus \, \textbf{3}_{(+1;+1)}  \, \oplus \, \textbf{3}^{\prime}_{(+1;-1)} \\
\, \oplus \, \textbf{3}_{(-\frac{3}{2};-\frac{1}{2})} \, \oplus \, \textbf{3}^{\prime}_{(+\frac{7}{2};+\frac{1}{2})}  \, \oplus \, \textbf{3}^{\prime}_{(-\frac{3}{2};-\frac{5}{2})} \, \oplus \,
\underline{\textbf{6}^{\prime}_{(-\frac{3}{2};-\frac{1}{2})}}  \, \oplus \, \textbf{8}_{(-\frac{3}{2};+\frac{3}{2})}\end{array}\right\} & ,
\end{array} \notag
\ee
where the underlined irreps are the only ones having the correct $\Sigma$ and $\Phi$ scaling weights to be interpreted as $h$, $f_{0}$ and $\omega$ fluxes respectively. The resulting dictionary is given 
in \eqref{dictionary_fluxes_max}.

By plugging \eqref{dictionary_fluxes_max} into the QC given in \eqref{QC_max}, one finds that $h \, f_{0} \, = \, 0$ should be imposed on the gauge fluxes, whereas no condition on $\omega$ needs to be 
imposed for consistency. This is in agreement with what predicted by dimensional reduction, where $h \, f_{0} \, = \, 0$ is interpreted as the absence of branes which is necessary for maximal supersymmetry
and where the only constraints on $\omega$, which are of the form
\be
{\omega_{[ij}}^{i'} \, {\omega_{k]i'}}^{l} \, = \, 0 \ ,
\ee
are automatically satisfied by any symmetric $q^{ij}$.

\section{Type IIA supergravity on the $\mathbb{T}^3/\mathbb{Z}_2$ orientifold}\label{IIACOMP}
We consider the internal space $\mathbb{T}^3/\mathbb{Z}_2$ with a spacetime-filling O6 source that sits at the eight fixed points of the $\mathbb{Z}_2$ involution:
\begin{equation}
(a, b, c)\ ,\qquad \text{where}\qquad a, b, c \in \{0, \tfrac{1}{2} \}\ .
\end{equation} 
The seven-dimensional gauged supergravity is a no-scale supergravity whose vacuum is non-supersymmetric \cite{Blaback:2012mu}. This supergravity, obtained from direct dimensional reduction, captures a consistent subset of the ten-dimensional fields that are capable of describing spontaneous supersymmetry-breaking in the vacuum. 

For the purpose of constructing vacua and domain wall solutions we used a truncation of this theory down to the metric and two scalar fields ($\rho$ and $\tau$), which was originally described in \cite{Blaback:2012mu}. But for completeness we briefly describe the full dimensional reduction of the bosonic sector revealing the presence of all the bosonic seven-dimensional fields.

Taking into account the parity rules for O6-planes ($B$ and $C_1$ are odd, $C_3$ is even) we have the following bosonic field content in seven dimensions: 1 metric field, $10  + 3N$  scalars, $6+ N$ vectors and 1 three-form.\footnote{Note that a massless three-form in seven dimensions is dual to a massless two-form.} The (unwarped) effective theory in $D= 7$ should be a half-maximal gauged supergravity coupled to $N$ vector multiplets. The scalar coset is
\begin{equation}
\mathbb{R}^{+}\times\frac{\textrm{SO}(3, 3+N)}{\textrm{SO}(3)\times \textrm{SO}(3+N)}\ .
\end{equation}
The number $N$ equals the number of D6-branes in the compact manifold. If we denote the flux quanta of $H$ and $F_0$  by the integers $n$ and $M$, then the allowed values for $N$ come from the tadpole condition
\begin{equation}\label{quant1}
nM = 16 - N \ .
\end{equation}
When the fluxes are turned off and we put sixteen D6-branes in the background of the eight O6-planes, they generate a $\textrm{U}(1)^{16}$ gauge group, when they are at different positions. In what follows we take $N=0$ and perform the explicit dimensional reduction of the ten-dimensional action.

The ten-dimensional action in Einstein frame is given by the sum of a bulk kinetic action, a WZ piece, and a local action, describing the O6/D6 configuration. The bulk kinetic term is given by
\begin{equation}
S =  \int \left(\star R -\tfrac{1}{2}\star\d\phi\wedge\d\phi -\tfrac{1}{2}\e^{-\phi}\star H\wedge H -\tfrac{1}{2}\sum_{n=0,2,4 }\e^{\tfrac{5-n}{2}\phi}\star F_n\wedge F_n \right)  \ ,
\end{equation}
where $H=\d B$, $m=F_0$ is the Romans mass and
\begin{align}
& F_2 = \d C_1 + m\,B\ ,\\
& F_4 = \d C_3 -H\wedge C_1 + \tfrac{1}{2}m B\wedge B\ .
\end{align}
The Wess-Zumino piece reads:
\begin{equation}
S =\int \left(+\tfrac{1}{40} m^2 B\wedge B \wedge B\wedge B\wedge B +\tfrac{1}{6}m B\wedge B\wedge B\wedge\d C_3 +\tfrac{1}{2} \d C_3\wedge\d C_3\wedge B \right)\ .
\end{equation}
The source term for O6-planes is
\begin{equation}
S = -T_{6} \, \e^{\tfrac{3}{4}\phi}\int_7\sqrt{-g_7} \,+\, T_{6} \int C_7\ .
\end{equation}

Now we will reduce on $\mathbb{T}^3/\mathbb{Z}_2$, taking into account that $B$ is odd, $C_1$ is odd and $C_3$ is even. There are no KK vectors due to the involution symmetries. Instead of using $\tau$ and $\rho$ we use canonically normalized scalar fields $\phi$ and $\varphi$. In Einstein frame, the ten-dimensional Ansatz is given by
\begin{align}
&  \d s^2_{10} = \e^{2\alpha\varphi} \d s_7^2 + \e^{2\beta\varphi}M_{ab}\d x^a\otimes\d x^b\ , \\
& \hat{B} = B^{(1)}_a\wedge \d x^a\ , \\
& \hat{H}=\d \hat{B} + h\,\d x^1\wedge\d x^2\wedge\d x^3\ ,\\
& \hat{C}_1 = C^{(0)}_a\d x^a\ ,\\
& \hat{C}_3 = C^{(3)} + \tfrac{1}{2}C^{(1)}_{ab}\wedge \d x^a\wedge\d x^b \ ,\\
&  F_0 = m\ ,
\end{align}
with $\alpha=\frac{1}{4}\sqrt{\frac{3}{5}}$, $\beta=-\frac{5}{3}\alpha$  and $h$ is the $H$ flux quantum. Tadpole cancellation requires to take 
\begin{equation}
T_{6} = -2 |h m|\ .
\end{equation}
The translation between the canonically normalized scalars $\varphi, \phi$ and $\rho,\tau$  is as follows:
\begin{equation}\label{scalarredef}
\tau^{-2} = \e^{\phi/2 +2\alpha\varphi}\ ,\qquad \rho = \e^{\phi/2 -\frac{10\alpha }{3}\varphi}\ .
\end{equation}

In total there are $10$ scalars ($\varphi$, $\phi$, $M_{ab}$, $C^{(0)}_a$), 6 one-forms $B^{(1)}_a, C^{(1)}_{ab}$ and 1 three-form $C^{(3)}$ in seven dimensions. Reducing the $H, F_2 , F_4$ and $F_0$ terms gives:
\begin{align}
& \int_{10} -\tfrac{1}{2}\e^{-\phi}\star H \wedge H =  \int_7 -\tfrac{1}{2}\e^{-\phi}\Bigl( \e^{\tfrac{4}{3}\alpha\varphi}g^{ab}\star_7\d B^{(1)}_a\wedge \d B^{(1)}_b   + h^2 |g|^{-1}\e^{12\alpha\varphi}\star_7 1\Bigr)\ ,\\
& \int_{10} -\tfrac{1}{2}\e^{\tfrac{5}{2}\phi}\star F_0 \wedge F_0 = \int_7 -\tfrac{1}{2}\e^{\tfrac{5}{2}\phi}\e^{2\alpha\varphi} m^2 \star_7 1\ ,\\
& \int_{10} -\tfrac{1}{2}\e^{\tfrac{3}{2}\phi}\star F_2 \wedge F_2 = \int_7 -\tfrac{1}{2}\e^{\tfrac{3}{2}\phi}\e^{\tfrac{10}{3}\alpha\varphi}\Bigl(g^{ab}\star_7 D C_a^{(0)}\wedge D C_b^{(0)} \Bigr)\ ,\\
& \int_{10} -\tfrac{1}{2}\e^{\tfrac{1}{2}\phi}\star F_4 \wedge F_4 = \int_7 -\tfrac{1}{2}\e^{\tfrac{1}{2}\phi}\Bigl(\e^{-6\alpha\varphi}\star_7\d C^{(3)}\wedge\d C^{(3)} +2 \e^{\tfrac{14}{3}\alpha\varphi}\star_7 S_{ab}\wedge S^{ab}  \Bigr)\ ,
\end{align}
where, from now on, $\star_7$ is the hodge star operator defined by the Einstein frame metric $\d s_7^2$ and $g_{ab}$ is the three-dimensional internal metric $\d s_3^2$. We furthermore defined 
\begin{align}
& DC_a^{(0)} = \d C_a^{(0)} + m B^{(1)}_a\ ,\\
&S_{ab} =  \tfrac{1}{2}\d C^{(1)}_{ab} + C^{(0)}_{[a}\d B^{(1)}_{b]} -\tfrac{1}{2}m B^{(1)}_a\wedge B^{(1)}_b\ ,\\
&S^{ab}= g^{ac}g^{bd}S_{cd}\ , 
\end{align}
where the first expression is a covariant derivative \`a la St\"uckelberg.

The reduction of the Einstein--Hilbert term gives kinetic terms for the $M_{ab}$ fields and the $\varphi$ field:
\begin{equation}
\int_7 \left( -\tfrac{1}{2}\star_7 \d\varphi\wedge\varphi +\tfrac{1}{4}\star_7\d M_{ab}\wedge \d M^{ab} \right) \ .
\end{equation}
$M^{ab}$ is the inverse of the $M_{ab}$ matrix: $M_{ab}M^{bc}=\delta^c_a$. 
Finally, the reduction of the WZ term gives
\begin{equation}
\int_7 \left( m \d C^{(3)}\wedge B_1^{(1)}\wedge B_2^{(1)}\wedge B_3^{(1)} + \tfrac{1}{2}\varepsilon^{abc}\d C^{(3)}\wedge  C^{(1)}_{ab}\wedge B_c^{(1)}\right) \ ,
\end{equation}
where $\varepsilon^{abc}$ is the totally antisymmetric symbol (not tensor). 

The only contribution to the scalar potential in $D=7$ comes from the Romans mass term, the tension\footnote{The reduction of the tension term gives $\int_7  2|hm|\e^{\tfrac{3}{4}\phi + 7\alpha \varphi} \star_7 1$.} and the $H$ flux: 
\begin{equation}
V = \tfrac{1}{2}\Bigl( |h|\e^{-\tfrac{1}{2}\phi + 6\alpha\varphi} - |m|\e^{\tfrac{5}{4}\phi+\alpha\varphi}\Bigr)^2\ .
\end{equation}
This coincides with the previous expression (\ref{V_Fluxes_half}) upon the identification of the scalars given in (\ref{scalarredef}).

The solutions to this seven-dimensional theory are in one to one correspondence with the ten-dimensional equations of motion with smeared O6-planes. This can be explicitly checked and a simple way to understand this is by noting that we have employed a consistent dimensional reduction scheme which by definition maps solutions of the lower-dimensional action to solutions of the higher-dimensional theory. But since we integrated over the internal manifold, switched off warping, assumed no internal coordinate dependence in the dilaton and no $F_2$ profile, we solved those ten-dimensional equations for which delta sources are  replaced by constant finite numbers \cite{Blaback:2010sj}. \newpage

\bibliography{refs}
\bibliographystyle{utphysmodb}
\end{document}